\documentstyle[11pt]{article}
\font \msb=msbm10 scaled \magstep1
\newcommand{\bR}{\mbox{\msb R} }
\newcommand{\bC}{\mbox{\msb C} }
\newcommand{\bZ}{\mbox{\msb Z} }

\font \eul=eufm10 scaled \magstep2

\newcommand{\ggl }{\mbox{\eul g}}
\newcommand{\gh}{\mbox{\eul h}}
\newcommand{\gn}{\mbox{\eul n}}
\newcommand{\gb}{\mbox{\eul b}}

\def\a{{\alpha}}
\def\b{{\beta}}

\def\d{{\delta}}
\def\l{{\lambda}}
\def\m{{\mu}}
\def\s{{\sigma}}
\def\D{{\Delta}}
\def\bD{{\bf\Delta}}
\def\te#1{{\widetilde{#1}}}
\newcommand{\gc}{\mbox{\bf I}}

\def\SL{{\cal L}}

\def\tr{{\rm tr}}
\def\nn{ \nonumber }
\def\bq{ \begin{equation} }
\def\eq{ \end{equation} }
\def\ben{ \begin{eqnarray} }
\def\en{ \end{eqnarray} }
\def\ll{ \label }
\def\dfrac#1#2{{\displaystyle{#1\over#2}}}
\def\frac#1#2{{{#1\over#2}}}

\begin{document}
\title{On superintegrable systems closed to geodesic motion.}
\author{
 A.V. Tsiganov\\
{\small\it
 Department of Mathematical and Computational Physics,
 Institute of Physics,}\\
{\small\it
St.Petersburg University,
198 904,  St.Petersburg,  Russia}
}
\date{}
\maketitle

\begin{abstract}
In this work we consider superintegrable systems in the
classical $r$-matrix method. By using other authomorphisms
of the loop algebras we construct new superintegrable systems
with rational potentials from geodesic motion on $\bR^{2n}$.
\end{abstract}
\vskip 0.5cm

\section{Introduction}
\setcounter{equation}{0}
We shall consider classical integrable hamiltonian systems  on
the coadjoint orbits of finite-dimensional Lie algebras
according to \cite{arn89}.  The dual space $\ggl^*$ to the Lie
algebra $\ggl$ is equipped with the natural Lie-Poisson
brackets specified by the condition that the Poisson bracket of
two linear functions on $\ggl^*$ coincides with their Lie
bracket in $\ggl$. Let $H$ be a function on $\ggl^*$, $\nabla
H\in \ggl$ the gradient of $H$.  In the space
$C^\infty(\ggl^*)$ of smooth function $H$ determines the
evolution with the associated hamiltonian equation
\bq
\dot{x}=-({\rm ad}^*_{\nabla H})\cdot x\,,\qquad
x\in \ggl^*\,.\ll{heq1}
\eq
If $\ggl$ is self-dual, i.e. has a nondegenerate inner product
which allows to identify $\ggl^*$ with $\ggl$ and ${\rm ad}^*$
with ad, then (\ref{heq1}) takes on the usual form
\bq
\dot{x}=\{H,x\}=
-({\rm ad}_{\nabla H})\cdot x\,,\qquad
x\in \ggl\,.\ll{heq2}
\eq
Henceforth we shall always to identify $\ggl^*$ with $\ggl$
and ${\rm ad}^*$ with ad.

Function $I\in C^\infty(\ggl)$ is called an integral of
evolution with a hamiltonian $H$ if
\[\{H,I\}=0\,.\]
The evolution on a $2n$-dimensional symplectic manifold $M$
with the hamiltonian $H$ is called completely integrable if
there exists $n$ functions $I_1,\ldots,I_n$, which are
independent integrals in the involution for the hamiltonian
$H$
\bq
\{I_i,I_j\}=0\,,\qquad i,j\leq n\,.
\ll{inv}
\eq
The functions $I_1,\ldots,I_n$ are independent, if forms
$dI_1,\ldots,dI_n$ are linearly independent on the common level
surfaces of these functions.

The evolution on a  manifold $M$, ${\rm dim}\,M=2n$ with the
hamiltonian $H$ is called superintegrable or degenerate, if
there exists more than $n$ independent integrals of motion
$\{I_j\}_{j=1}^k\,$, $k>n$ and $n$ of which are in the
involution (\ref{inv}) \cite{arn89,msvw67,sw66}.  For the
superintegrable systems all the integrals $\{I_j\}_{j=1}^k\,,$
$k>n$ are generators of the polynomial associative algebra,
whose defining relations are polynomials of certain order in
generators (see \cite{bb88,vin95} for a collection of original
papers).  The main example of the superintegrable systems is a
free motion with the following hamiltonian and equations of
motion

\bq H=\sum_{j=1}^n p_j^2\,,\qquad \dot{q_j}=p_j\,,
\qquad \dot{p_j}=0\,,\ll{fmot}
\eq
here $p_j,q_j$ are canonical variables on $M=\bR^{2n}$.
Integrals of motion in the involution and
additional integrals of motion may be defined as
\bq I_k=f(p_1,\ldots,p_n)\,,\qquad I_{jk}=p_jq_k-q_jp_k\,.
\ll{fint}\eq
Other known classical superintegrable systems with an arbitrary
numbers of degrees of freedom are the harmonic oscillator, the
Kepler problem and the Calogero system \cite{pe91}. Notice, that
the Kepler problem and the Calogero model may be obtained from the
geodesic motion (\ref{fmot}) on spaces of constant curvature
\cite{pe91}.  Another examples of superintegrable systems can
be constructed by using either purely algebraic techniques
\cite{arn89, bb88, vin95, msvw67} or separation of variables
method at $n=2,3$ \cite{sw66,ev90,vin95}. Some individual
examples of superintegrable systems are listed in \cite{pe91}
with the corresponding references.

Our aim is to show how superintegrable systems fit into a
general pattern based on the notion of the classical $r$-matrix
\cite{ft87,rs87}.  The main advantage of such embedding is that
important structure elements for superintegrable systems, such
as a Lax representation, separation of variables \cite{vin95}
and spectrum-generating algebra (dynamical algebra)
\cite{bb88,mm79} can be systematically derived from the
underlying standard $r$-matrix formalism, which is a
prerequisite for the study of the quantum case.

We propose a dressing procedure allowing to construct the new
superintegrable systems starting from known ones. As a natural
initial point we shall select a geodesic motion (\ref{fmot}) on
the Riemannian spaces of constant curvature.  The Lax
representations for these superintegrable geodesic motion are
known  \cite{kuz92}.  To construct the new Lax equations
associated to a potential superintegrable motion we apply the
outer automorphism of the corresponding loop algebras
\cite{ts96b} directly to the Lax equations associated to a
geodesic motion.

The paper is organized as follows.  In Section 2 we briefly
recall notion of classical $r$-matrix method.  By use a
triangular decomposition of semi-simple Lie algebras, in
Section 3, the algebraic approach to superintegrable systems is
proposed and this scheme is applied to several
examples.  In Section 4 superintegrable systems are constructed
in $r$-matrix formalism, while Section 5 contains some
examples.


\section{Method of the classical $r$-matrix}
\setcounter{equation}{0}
A systematic way for realizing integrable hamiltonian system on
coadjoint orbits of the Lie algebras is provided by the
$r$-matrix method \cite{ft87,rs87}.

Recall that the classical $r$-matrix on
a Lie algebra $\ggl$ is a linear operator $R\in {\rm End}(\ggl)$
such that the bracket on $\ggl$
\bq
\left[X,Y\right]_R=\dfrac12 \left(
[RX,Y]+[X,RY]\right)\,,\qquad X,Y\in\ggl\,,\ll{rbr}
\eq
satisfies the Jacobi identity \cite{rs87,rs88}. In this case
there are two structures of a Lie algebra on the linear space $\ggl$
given by original Lie bracket and by the
$r$-bracket (\ref{rbr}), respectively.  The Casimir functions $\tau_j$
on $\ggl^*$ invariant with respect to the original Lie structure
are in the involution with respect to the $r$-bracket.
If $\tau$ is an invariant function on $\ggl^*$, the associated
hamiltonian equation (\ref{heq1}) on $\ggl^*$ is equal to
\bq
\dfrac{dL}{dt}=-{\rm ad}^*_A\cdot L\,,\qquad
A=\dfrac12 R(d\tau(L))\,,\qquad L\in\ggl^*\,.\ll{lax1}
\eq
If $\ggl$ is self-dual, then (\ref{lax1}) takes on the usual
Lax form \cite{rs87}.

It is obvious, that for any $r$-matrix in (\ref{lax1}) all the
Casimir functions give rise to integrals of motion in the
involution \cite{ft87,rs87}. We have to find the origin of an
appearance of the  special superintegrable hamiltonians and
their additional integrals of motion.  Application of the
ad-invariant functions $\tau_j$ is a basic tool in the
$r$-matrix method \cite{ft87,rs87} and, therefore, we consider
these functions in greater detail to assume the standard
identification of the dual spaces.

To begin with let us recall some necessary facts from the
notion of a universal enveloping algebra
\cite{dix74}. Let $\ggl$ be a Lie algebra and $T(\ggl)$ be
the tensor algebra of the vector space $\ggl$
\bq
T=T^0\oplus T^1\oplus T^2\ldots\,,\qquad
T^n=\ggl\otimes\ggl\otimes\ldots\otimes\ggl\quad
n~{\rm times}\,.
\ll{talg}
\eq
If $J$ be the two-sided
ideal of $T$ generated by the tensors
\[x\otimes y-y\otimes x -[x,y]\,,\qquad x,y\in\ggl\,,\]
then the associative algebra $T/J$ is termed the universal
enveloping algebra, which is usual denoted by $U(\ggl)$.

Let $m\geq 0$ be an integer. The vector subspace of $U(\ggl)$
generated by the products $x_1x_2\cdots x_j$, where
$x_1,x_2,\ldots,x_j\in\ggl$ and $j\leq m$ is denoted by
$U_m(\ggl)$.  We have
\[U_0(\ggl)=\bC\cdot1\,,\qquad U_1(\ggl)=\bC\cdot 1
\oplus \ggl\,,\qquad
U_i(\ggl)U_j(\ggl)\subset U_{i+j}(\ggl)\,.\]
This sequence is termed the canonical filtration of $U(\ggl)$.

According by the Birkhoff-Witt theorem $T(\ggl)=J\oplus
S(\ggl)$ and algebra $U(\ggl)$  is isomorphic to the symmetric
algebra $S(\ggl)$ as a vector space.
If $x_1,x_2,\ldots,x_m\in\ggl$, then
\bq
w(x_1x_2\cdots x_m)=
\dfrac{1}{m!}\sum_\pi P_\pi x_1x_2\cdots x_m=
\dfrac{1}{m!}\sum_\pi x_{\pi(1)}x_{\pi(2)}\cdots x_{\pi(m)}\,,
\ll{symm}
\eq
here $P_\pi$ means the permutation operator corresponding to a
certain Young diagram $\pi$ \cite{dix74}.  The map $w$
(\ref{symm}) is a bijection of $S(\ggl)$ onto $U(\ggl)$, which
is called the symmetrization.

The Casimir functions or the ad-invariant functions on $\ggl$
form a center $\gc(\ggl)$ of $U(\ggl)$.  The symmetrization
mapping (\ref{symm}) allows us to construct $\gc(\ggl)$ by
using the ad-invariants of commutative algebra $S(\ggl)$
\cite{dix74}.

The most interesting class of examples in the classical
$r$-matrix method is provided by loop algebra $\SL(\ggl,\l)$
\cite{rs87}.  Algebra $\SL(\ggl,\l)$ can be realized as an algebra
of the Laurent polynomials with coefficients in $\ggl$
\[\SL(\ggl,\l)=\ggl[\l,\l^{-1}]=
\left\{x(\l)=\sum_i x \l^i\,,\quad x\in\ggl\right\}\]
and with the commutator $[x\l^i,y\l^j]=[x,y]\l^{i+j}$.

According to an algebra homomorphism \cite{kac90}
\[U(\SL(\ggl,\l))\to \bC[\l,\l^{-1}]\otimes U(\ggl)\]
the Casimir functions on the loop algebras can be recovered
by the ad-invariants $\tau_j(x)$ in $\gc(\ggl)$
\bq
\tau_{j,\phi}=\left.{\rm Res}\right|_{\l=0}\phi(\l)\cdot\tau_j(x(\l))\,,
\quad \tau_j(x)\in\gc(\ggl)\,,
\quad \phi(\l)\in \bC[\l,\l^{-1}]\,,\ll{adinv}
\eq
where $\phi(\l)$ is some rational function on spectral
parameter $\l$ with numerical values \cite{rs87}.

Studying the superintegrable systems we have to introduce the
usual tensor algebra
\ben
T(\ggl,\l,\m,\ldots)&=&T^0\oplus T^1\oplus T^2\ldots\,,\nn\\
T^m(\ggl,\l,\m,\ldots,\nu)
&=&\SL(\ggl,\l)
\otimes\SL(\ggl,\m)\otimes\ldots\otimes\SL(\ggl,\nu)\quad-\quad
m-{\rm times}\,,
\ll{talgl}
\en
and canonical filtration of the corresponding enveloping
algebra $U(\ggl,\l,\m,\ldots)$ generated by subspaces
$U_m(\ggl,\l,\m,\ldots,\nu)$.  These vector subspaces are
produced by subspaces $U_m(\ggl)$
\[
x_{i_1i_2\cdots i_k}(\l,\m,\ldots,\nu)=
\sum_{j_1,j_2,\ldots,j_k}
x_{i_1i_2\cdots i_k}\l^{j_1}\m^{j_2}\cdots\nu^{j_k}\,,
\qquad k\leq m\,,\qquad x_{i_1i_2\cdots i_k}\in U_m(\ggl)\,.
\]
In just the same way as for a Lie algebra $\ggl$ \cite{dix74},
we can define the canonical mapping of the loop algebra
$\SL(\ggl,\l)$ into $U(\ggl,\l,\m,\ldots)$. Any element
$L(\l)$ of $\SL(\ggl,\l)$ can be embedded into
$U_m(\ggl,\l,\m,\ldots,\nu)$
\bq
L_j(\l_j)=id_1\otimes \cdots \otimes id_{j-1}\otimes L(\l_j)
\otimes id_{j+1}\otimes
\cdots\otimes id_m\in U_m(\ggl,\l,\m,\ldots,\nu)\,.\ll{emb1}\eq
and
\ben
&&L^{(k)}_{j_1j_2\cdots j_k}
(\l_1,\l_2,\ldots,\l_k)=\prod_{n=1}^k L_{j_n}(\l_{j_n})\,,
\qquad 1\leq k\leq m\ll{emb2}\\
&&L^{(m)}(\l,\m,\ldots,\nu)=L(\l)\otimes L(\m)\otimes\cdots
\otimes L(\nu)\,,\quad-\quad m-{\rm times}\,.\ll{emb3}
\en
here $\l_1=\l\,,\quad\l_2=\m,\quad\ldots\quad\l_m=\nu$.

If $\ggl$ is identified with its dual, then $r$-bracket (\ref{rbr})
can be rewritten in the tensor form \cite{ft87,rs87,rs88}.
Let $L(\l)\in\SL(\ggl,\l)$ be a generic point in the loop
algebra, which is regarded as a Lax matrix.  The corresponding
$r$-bracket is given by
\ben
\{L_1(\l),L_2(\m)\}&=&
[r_{12}(\l,\m),L_1(\l)]-[r_{21}(\l,\m),L_2(\m)\,]\,,\nn\\
\ll{rpoi}\\
r_{21}(\l,\m)&=&P_{12} r_{12}(\l,\m) P_{12}\,,\nn
\en
where $P_{12}$ is a permutation operator in
$\SL(\ggl,\l)\otimes \SL(\ggl,\m)$ and $r_{ij}(\l,\m)$ are
kernels of the corresponding operators $R$ and $R^*$ in
(\ref{rbr}) \cite{rs88}. Notice, that the $r$-matrix scheme is
extended easily to the twisted subalgebras of loop algebra
$\SL(\ggl,\l)$ and the corresponding matrices $r_{12}$ have
rational, trigonometric and elliptic dependence on spectral
parameter.

The  Poisson brackets between the elements
$L^{(k)}_{j_1j_2\cdots j_k}(\l,\m,\ldots,\nu)$ (\ref{emb2})
can be written in the "generalized" $r$-matrix form. For instance
\ben
\{L_{12}(\l,\m),L_3(\nu)\}
&=&[r_{13}(\l,\nu)+r_{23}(\m,\nu),L_{12}(\l,\mu)]\nn\\
\ll{rp1}\\
&-&[r_{31}(\l,\nu),L_{23}(\m,\nu)\,]
-[r_{23}(\m,\nu),L_{13}(\l,\nu)\,]\,,\nn
\en
where $r_{ij}(\l_i,\l_j)$ are $r$-matrices, which act nontrivially in
the corresponding subspaces of $T_m(\ggl,\l,\m,\ldots,\nu)$
(\ref{talgl}) and
\ben
\{L_{12}(\l,\m),L_{34}(\nu,\eta)\}
&=&\left[r^{(1)}(\l,\m,\nu,\eta),L_1(\l)\right]
+\left[r^{(2)}(\l,\m,\nu,\eta),L_2(\m)\right]\nn\\
\ll{rp2}\\
&-&\left[r^{(3)}(\l,\m,\nu,\eta),L_3(\nu)\right]
-\left[r^{(4)}(\l,\m,\nu,\eta),L_4(\eta)\right]\,,\nn
\en
where
\ben
r^{(1)}(\l,\m,\nu,\eta)&=&\te{r}(\l,\m,\nu,\eta)+
P_{34}\te{r}(\l,\m,\eta,\nu) P_{34}\,,\nn\\
r^{(2)}(\l,\m,\nu,\eta)&=&P_{12}r^{(1)}(\m,\l,\nu,\eta)P_{12}\,,\nn\\
r^{(3)}(\l,\m,\nu,\eta)&=&P_{13}
\left[\te{r}(\l,\m,\nu,\eta)+
P_{12} \te{r}(\m,\l,\nu,\eta) P_{12}\right]
P_{13}\,,\nn\\
r^{(4)}(\l,\m,\nu,\eta)&=
&P_{34}r^{(3)}(\l,\mu,\eta,\nu)P_{34}\,,\nn\\
\te{r}(\l,\m,\nu,\eta)&=&r_{13}(\l,\nu)L_{24}(\m,\eta)\,.\nn
\en
Here $P_{ij}$ are operators of pairwise permutations in the
tensor algebra $T_m(\ggl,\l,\m,\ldots,\nu)$ (\ref{talgl}).

Integrals of motion in the involution  are completely defined
by the ad-invariants (\ref{adinv}) of $\SL(\ggl,\l)$ and by the
linear $r$-bracket (\ref{rpoi}) \cite{ft87,rs87}. For
description of the superintegrable systems we have to consider
ad-invariants of $U(\ggl,\l,\m,\ldots)$ and more complicated
embedding $L^{(k)}_{j_1j_2\cdots j_k}(\l,\m,\ldots,\nu)$
(\ref{emb2}-\ref{emb3}).  In the next section we consider
several examples, that allows us to understand the origin of
the appearance of additional integrals of motion in the
classical $r$-matrix method.

\section{One class of superintegrable systems}
\setcounter{equation}{0}
Let $\ggl$ be a simple Lie algebra with a
Cartan subalgebra $\gb$ and a system of simple roots
$\Pi=\{\a_1,\a_2,\ldots,\a_n\}$. Let
\[e_i=e_{\a_i}\,,\qquad h_i=h_{\a_i}\,,\qquad f_i=e_{-\a_i}\,,\qquad
i=1\,\ldots n\,,\]
be the Chevalley generators and $\{h_i,e_i,f_i\}$ be
a Cartan-Weil basis \cite{bour68}, normalized by
$(e_i,f_i)=1$:
\ben
&[h_i,h_j]=0\,,\qquad&[e_i,f_j]=\d_{ij}h_i\,,\nn\\
&[h_i,e_j]=c_{ij}e_j\,,\qquad& [h_i,f_j]=-c_{ij}f_j\,,\ll{basis}\\
&({\rm ad}e_i)^{-c_{ij}+1}\cdot e_j=0\,,\qquad&
({\rm ad}f_i)^{-c_{ij}+1}\cdot f_j=0\,.\nn
\en
Here $c_{ij}$ are entries of the Cartan matrix, which are integers
in this normalization.

Let $\gn_+$ ($\gn_-$) be the linear span of root vectors
$e_j$ ($f_j$) such that
\bq
\ggl=\gn_-\oplus\gb\oplus\gn_+\,.\ll{trdec}\eq
This decomposition is termed the triangular decomposition of
$\ggl$.

Every basis in the commutative subalgebra $U_m(\gb)\subset
U_m(\ggl)$ could be associated to the family of functionally
independent integrals in the involution.  Let us define a
hamiltonian $H$ as a function on alone generator $h_i$ of the
Cartan subalgebra $\gb$
\bq H=H(h_i)\,,\qquad h_i\in\gb\,,
\qquad H\in U_m(\gb)\,.\ll{sham}\eq

It is immediately seen that the $n-1$ functionally independent
integrals in the involution may be constructed for the
hamiltonian $H$ (\ref{sham}) in $U_m(\gb)$
\bq I_j=I_j(h_1,\ldots,h_n)\,,\qquad h_i\in\gb\,,\qquad
I_j\in U_m(\gb)\,,\ll{sint1}\eq
here $n={\rm dim}\gb={\rm rank} \ggl$.

Moreover, we can introduce another $n(n-1)$ integrals of evolution
in $U(\gn_-\oplus\gn_+)$
\ben
&&I_{jk}^{lm}=e_j^l\cdot f_k^m\in U_p(\gn_-\oplus\gn_+)\,,\quad
\dfrac{l}{m}=\dfrac{c_{ik}}{c_{ij}}\,, \quad p=\max\{l,m\}\,, \nn\\
\ll{adint}\\
&&\{ H(h_i),I_{jk}^{lm} \}=(lc_{ij}-mc_{ik})
\dfrac{dH}{dh_i}\cdot I_{jk}^{lm}=0\,,\nn
\en
where $l\geq 1$ and $m\geq 1$ in monomials $I_{jk}^{lm}$
are positive integers in normalization (\ref{basis}).

Let $p\geq 1$ be a largest power of monomials $I_{jk}^{lm}\,,$
$j,k=1,\ldots,n$. Taking (\ref{basis}) into account we observe that
other monomials in $U_p(\ggl)$
\bq
T_{k_1k_2k_3}^{j_1j_2j_3}
=h_{k_1}^{j_1}e_{k_2}^{j_2}f_{k_3}^{j_3}\,,\qquad
j_1+j_2+j_3\leq p\,,\ll{gladd}
\eq
belong to the generalized dynamical algebra \cite{bb88,vin95}
defined as
\bq\{H(h_i),T_{k_1k_2k_3}^{j_1j_2j_3}\}=
\left[(j_2c_{ik_2}-j_3c_{ik_3})\dfrac{dH}{dh_i}\right]\cdot
T_{k_1k_2k_3}^{j_1j_2j_3}\,.\ll{gda}\eq
In this case the generalized dynamical algebra \cite{bb88,vin95}
of $H$ (\ref{sham}) generated by elements (\ref{gda})
coincides with the polynomial subalgebra $U_p(\ggl)$.
Operators
\bq Q_{k_1k_2k_3}^{j_1j_2j_3}=\left(\dfrac{dH}{dh_i}\right)^{-1}
T_{k_1k_2k_3}^{j_1j_2j_3}\,,\ll{ladd}\eq
are counterparts of the ladder operators of hamiltonian $H$
(\ref{sham}) in quantum mechanics.  If the hamiltonian $H$ is a
linear function on generators, the spectrum-generating algebra
\cite{bb88,mm79} generated by elements (\ref{ladd}) coincides
with the polynomial subalgebra $U_p(\ggl)$.

The realization of a proposed scheme for a given hamiltonian is
a quite difficult task, which is not always dictated by an a
priori obvious procedure of the searching of a necessary
algebra $\ggl$. Moreover, the different algebras $\ggl$ could
be associated to the single superintegrable system.

On the other hand, taking some fixed algebra
$\ggl$, we have to introduce a suitable representation for a
realization of superintegrable system in canonical variables on
it's coadjoint orbits.

Let us take the $n$
three-dimensional Heisenberg algebras $\gh_j$
($\ggl=\oplus \gh_j$) having basis
($h_j,e_j,f_j$) such that $[e_j,f_j]=h_j$ and
$[h_j,e_j]=[h_j,f_j]=0$.  The center of nilpotent algebra
$\gh_j$ is $\bC\cdot h_j$.
The corresponding universal enveloping algebra $U(\ggl=\oplus
\gh_j)$ is a Weil algebra $W_n$. Polynomial subalgebra
$U_2(\ggl)$ of $W_n$ is well-known in mathematics
\cite{arn89,dix74} and physics \cite{bb88,ev90,mm79}. It
consists of quadratic generators $T_{ij}$, $P_{ij}$, $S_{ij}$
and linear generators $(h_j,e_j,f_j)$
\bq
T_{ij}=e_if_j\,,\qquad S_{ij}=e_ie_j\,,\qquad
P_{ij}=f_if_j\,,\ll{ajk}
\eq
where we do not consider polynomials in elements of the center
of $\gh_j$.

The non-vanishing polynomial Poisson brackets in $U_2(\ggl)$ are
\ben
\{e_i,T_{jm}\}&=&h_i\d_{im}e_j\,,\qquad
\{e_i,P_{jm}\}=h_i\left(\d_{ij}f_m+\d_{im}f_j\right)\,,\nn\\
\{f_i,T_{jm}\}&=&-h_i\d_{ij}f_m\,,\qquad
\{f_i,S_{jm}\}=-h_i\left(\d_{ij}e_m+\d_{im}e_j\right)\,,\nn\\
\{T_{ij}\,,T_{lm}\,\}&=&
\left(\d_{im}h_i T_{lj} -\d_{jl}h_j T_{im} \,\right)\,,\ll{spalg}\\
\{T_{ij},S_{lm}\}&=&-h_j\left(\d_{jm}S_{il}+\d_{jl}S_{im}\right)\,,\qquad
\{T_{ij},P_{lm}\}=h_i\left(\d_{im}P_{jl}+\d_{il}P_{jm}\right)\,,\nn\\
\{S_{ij},P_{lm}\}&=&h_j\left(\d_{jm}T_{il}+\d_{jl}T_{im}\right)+
h_i\left(\d_{im}T_{jl}+\d_{il}T_{jm}\right)\,,\nn
\en
The coadjoint orbits of the Heisenberg algebras are planes
$h_j=\a_j\in\bC$.  On the fixed orbit of the algebra
$\ggl=\oplus \gh_j$ we can reinterpret the brackets (\ref{spalg})
not as polynomial brackets in generators $\ggl$, but rather
as linear brackets between generators $T_{ij}$,
$P_{ij}$, $S_{ij}$ and $e_j$, $f_j$ with the structure
constants $h_j=\a_j$. It allows us to identify representations of
$U_2(\ggl)$ with representation of the known Lie algebras.
If  we consider $n$ equivalent orbits of nilpotent algebra
$\gh$ ($\a_i=\a_j$), the complete algebra $U_2(\ggl)$ is
isomorphic to the semi-direct product $W_n\otimes sp(2n)$
\cite{bb88,ev90}.  Subalgebra of generators $T_{ij}$ is
isomorphic to the algebra $u(n)$ or, after separation of
Casimir operator $\sum T_{kk}$, to the algebra $su(n)$.

The $n$-dimensional commutative ideal in $U_2(\ggl)$ (the Cartan
subalgebra) is generated by elements $\te{h_j}=e_jf_j=T_{jj}$.
If we introduce the hamiltonian $H$ by the rule (\ref{sham}) on
this $n$-dimensional ideal, then another generators of
$U_2(\ggl)$ have the following brackets with it
\ben
&H=\sum_{i=1}^n \te{h}_j=\sum_{j=1}^n T_{jj}\,,
\qquad &\{H,T_{ij}\}=(h_i-h_j) T_{ij}\,,\nn\\
&\{H,e_j\}=-h_j e_j\,,\qquad &\{H,f_j\}=h_j f_j\,,\ll{dynalg}\\
&\{H,S_{ij}\}=-(h_i+h_j) S_{ij}\,,\qquad
&\{H,P_{ij}\}= (h_i+h_j) P_{ij}\,.\nn
\en
The hamiltonian $H$ is a superintegrable hamiltonian in
$U_2(\ggl)$ on the following coadjoint orbit of $\ggl=\oplus
\gh_j$
\bq h_i=h_j\,,\qquad i,j=1,\ldots,n\,.\ll{sorb}\eq
Note, that we can select different $n$-dimensional commutative
ideals in the various subalgebras $U_m(\ggl)$ to construct
superintegrable systems (\ref{sham}-\ref{adint}) with some
fixed hamiltonian (\ref{sham}).

The $n$-dimensional harmonic oscillator in the euclidean
coordinates is described by the hamiltonian
\bq
H=\dfrac12\sum_{j=1}^n \left( p_j^2+
\a_j^2 q_j^2\right)\,.\ll{h0}
\eq
The phase space of oscillator (\ref{h0}) can be identified with
the coadjoint orbits of Heisenberg algebras according to
\ben
e_j&=&a_j=\dfrac1{\sqrt2}(p_j-i\a_jq_j)\,,\nn\\
f_j&=&a_j^+=\dfrac1{\sqrt2}(p_j+i\a_jq_j)\,,\ll{halg}\\
h_j&=&i\a_j\,,\nn
\en
where $a_j,\quad a_j^+$ are the standard
creation and annihilation operators.
In this realization, on the orbit (\ref{sorb}) hamiltonian $H$
(\ref{h0}) of isotropic oscillator is superintegrable in $U_2(\ggl)$
\cite{arn89}.

However, it will be useful to identify the phase space of
oscillator with the coadjoint orbits of another algebras.  For
instance, if the phase space of oscillator be identified with
the coadjoint orbits in $sl(2)^*$, we can introduce integrable
systems isomorphic to oscillator
\bq
H'=\dfrac12\sum_{j=1}^n \left(p_j^2+\a_j^2q_j^2\right)+
\sum_{j=1}^n \dfrac{\b_j}{q_j^2}\,,\qquad \b_j\in\bR
\ll{hsw}
\eq
which is known as the Smorodinsky-Winternitz system at
$n=2,3$ \cite{sw66}.

Let us prove it.  Consider an infinite-dimensional
representation $W$ of the Lie algebra $sl(2)$ in linear space
$V$ defined in the Cartan-Weil basis $\{h,e,f\}$ in
$\rm{End}(V)$ equipped with the natural bracket
\bq [h,e]=e\,,\quad [h,f]=-f\,,\quad
[e,f]=2h\,,\quad
\bD=h^2+\dfrac12(ef+fe)\,.\ll{ssl2}
\eq
If operator $e$ is invertible in $\rm{End}(V)$, then the mapping
\ben
&&h\to h'=h\,,\qquad e\to e'=e\,, \nn\\
&&f\to f'=f+\b e^{-1}\,,\qquad \b\in\bC \ll{mapp}
\en
is an outer automorphism of the space of infinite-dimensional
representations of $sl(2)$ in $V$ \cite{ts96b}. The mapping
(\ref{mapp}) shifts spectrum of $\bD$ on the parameter $\b$
\[ \D \to \D'=\D+\b. \]
We can suppose that automorphisms (\ref{mapp}) defines
one-parametric realization $W(\b)$ of $sl(2)$.
For instance, realization of $sl(2)$ with one free parameter
$\b$ in the classical mechanics  is given by
\bq h=\dfrac{qp}2\,,\quad
e=\dfrac{q^2}2\,,\quad f=-\dfrac{p^2}2+\dfrac{\b}{q^2}\,,
\quad \D'=\b\,,\ll{clrep}
\eq
where $(q,p)$ is a pair of canonical coordinate and momenta with
the classical Poisson bracket $\{p,q\}=1$.
Motivated by realization (\ref{clrep}) we present one
application of the automorphism (\ref{mapp}) in the theory of
integrable systems.  Consider a classical hamiltonian system
completely integrable on the $\bR^{2n}$ with the
natural hamiltonian
\[H=\sum_{j=1}^n p_j^2+V(q_1,\ldots,q_n)\,.\]
Let the phase space be identified completely or partially with
the $m$ coadjoint orbits in $sl(2)^*$ as (\ref{clrep}).  Then
the mapping
\bq
H\to H'=H+\sum_{j=1}^m \dfrac{\b_j}{q_j^2}\,,\qquad \b_j\in\bR
\ll{pr1}
\eq
preserves the properties of integrability and separability.
The list of such systems can be found in \cite{pe91}.

Let us take the $n$ algebras $sl(2)$ ($\ggl=\oplus sl(2)$)
having basis ($h_j,e_j,f_j$) with the brackets
\bq [h_j,e_j]=\a_je_j\,,\quad [h_j,f_j]=-\a_jf_j\,,\quad
[e_j,f_j]=2\a_jh_j\,,\ll{tsl2}
\eq
where $\{\a_j\}_{j=1}^n\in\bC$ is a set of arbitrary constants.

As usual, we can define the polynomial subalgebra $U_2(\ggl)$
of the corresponding universal enveloping algebra.
Here we consider a special subalgebra of $U_2(\ggl)$ consisting
of quadratic operators $T_{ij}\in U_2(\gn_-\oplus\gn_+)\,,\quad
i\neq j$ and linear operators $T_{ii}\in U_1(\gb)$
\bq
T_{ij}=e_if_j\,,\qquad i\neq j\,,
\qquad
T_{ii}=h_i\,,\ll{tjk}\
\eq
These operators are closed under the Poisson bracket
\ben
\{T_{jk},T_{lm}\}&=&
2\left(\a_j\d_{jm}-\a_k\d_{lk}\right)T_{jm}T_{lk}\,,
\qquad j\neq k,\, l\neq m\,,\nn\\
\{T_{jj},T_{lm}\}&=&\a_j\left(\d_{jl}-\d_{jm}\right)T_{lm}\,,
\qquad l\neq m\,,\ll{sual}\\
\{T_{jj},T_{kk}\}&=&0\,,\nn
\en
or, in the unit polylinear form,
\ben
\{T_{jk},T_{lm}\}&=&
\left[ \a_j\d_{jk}\left(\d_{jl}-\d_{jm}\right)T_{lm}+
\a_l\d_{ml}\left(\d_{kl}-\d_{jl}\right)T_{jk}+  \right.\nn\\
&+&2\left.
(1-\d_{jk})(1-\d_{lm})
\left(\a_j\d_{jm}-\a_k\d_{lk}\right)T_{jm}T_{lk}\,,
\right]\,.\nn
\en
As before, we introduce linear
hamiltonian $H=\sum h_j=\sum T_{jj}$ in the Cartan subalgebra
of $\ggl=\oplus sl(2)$. Elements $T_{ij}$ of
$U_2(\gn_-\oplus\gn_+)$ have the following brackets with the
hamiltonian
\[\{H,T_{ij}\}=(\a_j-\a_i)T_{ij}\,.\]
The coadjoint orbits in $sl(2)$ algebras (\ref{tsl2}) are fixed
by the value of the Casimir operator $\bD$ (\ref{ssl2}):
$\D_j=\a_j^2\in\bC$.
On the special coadjoint orbit of $\ggl=\oplus sl(2)$
determined by $\a_i=\a_j\,,~i,j=1,\ldots,n$ the linear hamiltonian
$H$ (\ref{sham}) has additional integrals of evolution $T_{ij}$
(\ref{tjk}).

The phase space of oscillator can be identified with the
coadjoint orbit of $\ggl$ by using the representation
(\ref{clrep}) and the similar transformation
\bq
\left(\begin{array}{cc}h_j&e_j\\
f_j&-h_j\end{array}\right)=(\a_jU_j)^{-1}
\left(\begin{array}{cc}h_j'&e_j'\\
f_j'&-h_j'\end{array}\right)U_j\,,\qquad
U_j=\dfrac{1}{\sqrt{2}}
\left(\begin{array}{cc}i\a_j&-1\\
\a_j&-i\end{array}\right)\,.\ll{simtr}
\eq
Matrix elements $\{h_j',e_j',f_j'\}$ in (\ref{simtr})  are
generators of $sl(2)$ in representation (\ref{clrep}). More
explicitly
\ben
e_j&=&\dfrac14(p_j^2-\a_j^2q_j^2+2i\a_jp_jq_j)=\dfrac12(a_j^+)^2\,,\nn\\
f_j&=&\dfrac14(p_j^2-\a_j^2q_j^2-2i\a_jp_jq_j)=\dfrac12\,a_j^2\,,\nn\\
h_j&=&\dfrac{-i}4(p_j^2+\a_j^2q_j^2)=\dfrac{-i}2 N_j\,,
\ll{bbj}
\en
where $\{a_j,a_j^+,N_j\}_{j=1}^n$ are the standard operators (\ref{halg}).
Hamiltonian $H=2i\sum T_{jj}$ of oscillator  (\ref{h0}) or of
the Smorodinsky-Winternitz system (\ref{hsw}) and  integrals of
motion $T_{ij}$ (\ref{tjk}) are at most second order polynomials in
generators of $sl(2)^*$.

If the ratio of frequencies $\a_j/\a_k=l/m$ is rational, then
the hamiltonian of oscillator (\ref{h0}) has additional
integrals of motion in $U_p(\ggl)$ (\ref{adint}) \cite{arn89}.
The corresponding dynamical algebra coincides with $U_p(\ggl)$.
For instance, the anisotropic oscillator in two dimensions with
a $2:1$ ratio of the frequencies and with a third order
polynomial algebra $U_3(\ggl)$ has been considered in
\cite{vin95}.

So, we can see that for the superintegrable hamiltonian $H$
(\ref{sham}) in $U_m(\gb)$ the additional integrals of motion
belong to $U_m(\gn_-\oplus\gn_+)$. In the next section, motivated
by the presented examples, we consider superintegrable systems in the
classical $r$-matrix method.


\section{Superintegrable systems on $sl(n)$}
\setcounter{equation}{0}
In application to integrable systems every element $L(\l)$ of the loop
algebra $\SL(\ggl,\l)$ be a matrix in some fixed
matrix representation of the algebra $\ggl$ and representation
space be an auxiliary space \cite{ft87}.
Let us consider an algebra $\ggl=sl(n,\bC)$ in the fundamental
representation and begin with the standard rational $r$-matrix
\cite{ft87,rs87}.  In this case  element $L(\l)$ of $\SL(\ggl,\l)$ be
a $n\times n$ matrix in the auxiliary space $\bC^n$ and the matrix
elements of $L(\l)$ are some rational functions of spectral
parameter $\l$.

Basis of invariant functions in $\gc(sl(n,\bC))$ could be
selected as
\bq
\tau_k(L(\l))=\dfrac1k\tr L^k(\l)\,,\qquad k\leq n.\ll{tm}
\eq
The family of the integrals of motion
in the involution is generated by $\tau_k(L(\l))$ \cite{ft87,rs87}
\bq
I_{i,k}(L)=\Phi_{\l}^{(i)}(\tau_k(\l))\,,\ll{int}
\eq
where $\Phi_{\l}^{(i)}$ are various linear functionals defining
a set of the functionally independent integrals of
evolution.  For instance,
\bq
\Phi_{\l}^{(i)}(z)=\left.{\rm
Res}\right|_{\l=0}(\phi_i(\l)\cdot z)\,,
\qquad \phi(\l)\in \bC[\l,\l^{-1}]\,,
\ll{resfi}\eq
here $\phi_i(\l)$ are some functions of spectral parameter with
numerical values.  In this case integrals $I_{i,k}$ are at most
$k$-order polynomials in generators $\ggl$, $I_{i,k}\in
U_k(\ggl)$.

The Lax equation (\ref{lax1}) associated with the
hamiltonian $I_{i,k}$ (\ref{int}) is equal to
\bq
\dfrac{dL(\m)}{dt}=\left\{I_{i,k},L(\m)\right\}=
\left[L(\m),A_{i,k}(\m)\right]\,,\ll{laxmu}\eq
where the second Lax matrix $A_{i,k}$ has the form
\bq
A_{i,k}(\mu)=\Phi^{(i)}_{\l}
\tr_1\left(r_{21}(\l,\mu)L_1^{k-1}(\l)\right)\,, \ll{lax2}
\eq
here trace $\tr_1$ is taken over the first auxiliary space.

The representation space of the subalgebra
$U_m(\ggl,\l,\m,\ldots,\nu)$ be an extended auxiliary space
\bq
V^{(m)}=\otimes_{i=1}^m V_i=V_1\otimes V_2\otimes \cdots
\otimes V_m\,,\qquad V_i\simeq\bC^n\,.\ll{space}
\eq
The elements $L_j(\l_j)$ (\ref{emb1}) and $L^{(k)}_{j_1j_2\cdots
j_k}(\l,\m,\ldots,\nu)$ (\ref{emb2}) belonging to the algebra
$U_m(\ggl,\l,\mu,\ldots,\nu)$ are the $n^m\times
n^m$ matrices in $V^{(m)}$
\ben
&&L_j(\l_j)=I_1\otimes \cdots I_{j-1}\otimes L(\l_j)
\otimes I_{j+1}\otimes
\cdots\otimes I_m\,,\ll{embm}\ll{lj}\\
&&L^{(m)}(\l,\m,\ldots,\nu)=\prod_{j=1}^m L_j(\l_j)\,,\qquad
\l_1=\l\,,\quad\l_2=\m,\ldots,~\l_m=\nu\,,\nn
\en
here $I$ means a $n\times n$ unit matrix and subscript $j$
shows in which of the spaces $V_j$ in the whole space $V^{(m)}$
the matrix $L(\l)$ acts nontrivially.

The equation of evolution for the matrix $L^{(m)}$ has a
commutator Lax form
\bq
\dfrac{d}{dt}L^{(m)}(\l,\m\,\ldots,\nu)=
\left[L^{(m)}(\l,\m\,\ldots,\nu),A^{(m)}
(\l,\m\,\ldots,\nu)\right]\,,
\ll{laxm}
\eq
with the following second matrix
\bq
A^{(m)}(\l,\m\,\ldots,\nu)=\sum_{j=1}^m A_j(\l_j)\,,
\qquad\l_1=\l\,,\quad\l_2=\m,\ldots,\l_m=\nu \ll{m1}
\eq
which is a sum of the matrices $A_j(\l_j)$ of the type
(\ref{embm}) acting in the whole spaces $V^{(m)}$. The spectral
invariants of the matrices $L^{(m)}$ give rise to an involute
family of the integrals of motion as before.

Taking into account the symmetrization mapping $w$ (\ref{symm})
let us introduce matrices
\bq
L^{(m,\pi)}(\l,\m,\ldots,\nu)=P_\pi L^{(m)}(\l,\m,\ldots,\nu)=
P_\pi L_1(\l)L_2(\m)\cdots L_m(\nu)\,.
\ll{symmat}
\eq
Here permutation matrix $P_\pi$ in $V^{(m)}$ is determined by
\bq
P_\pi\left( x_1\otimes x_2\otimes\cdots\otimes x_m\right)=
x_{\pi(1)}\otimes x_{\pi(2)}\otimes\cdots\otimes x_{\pi(m)}\,,
\ll{prp}\eq
for any set of vectors $x_j$ in $\bC^n$ or
\bq
P_\pi\cdot A_1B_2\cdots D_m=A_{\pi(1)}B_{\pi(2)}\cdots
D_{\pi(m)}\cdot P_\pi\,,\qquad P_\pi^2=I\,,\ll{prp2}\eq
for any $n\times n$ matrices $A,B,\ldots,D$ embedding
in $V^{(m)}$ according to (\ref{embm}).  The permutation of
subscripts in (\ref{prp}) and (\ref{prp2}) is defined by a
certain Young diagram $\pi$ \cite{ks82}.

The equations of evolution for the matrices $L^{(m,\pi)}$ are
equal to
\bq
\dfrac{d}{dt}L^{(m,\pi)}(\l,\m\,\ldots,\nu)=
L^{(m,\pi)}A^{(m)}(\l,\m\,\ldots,\nu)-
A^{(m,\pi)}(\l,\m\,\ldots,\nu)L^{(m,\pi)}\,,
\ll{laxs}
\eq
where  matrix $A^{(m)}$ is given by (\ref{m1}) and the second
matrix $A^{(m,\pi)}$ differs from it by the permutation of
spectral parameters in accordance with the Young diagram $\pi$
\ben
A^{(m,\pi)}(\l,\m\,\ldots,\nu)
&=& P_\pi A^{(m,\pi)}(\l,\m\,\ldots,\nu)P^{-1}_\pi\,,\nn\\
&=&\sum_{j=1}^m A_j(\l_{\pi(j)})\,,
\quad\l_1=\l\,,\quad\l_2=\m,\ldots,\l_m=\nu \ll{m2}\nn
\en
The right-hand side of (\ref{laxs}) is a matrix commutator if
and only if
\bq
A^{(m)}(\l,\m\,\ldots,\nu)=A^{(m,\pi)}(\l,\m\,\ldots,\nu)\,,
\quad\Longleftrightarrow\quad
A(\l_j)=A(\l_{\pi(j)})\,.\ll{eq1}
\eq
Assuming (\ref{eq1}) holds we can define
new multivariable generating functions of the integrals of
motion
\ben
&&s_m^{\pi}(\l,\m\,\ldots,\nu)=
\dfrac1m \tr_{(m)}L^{(m,\pi)}(\l,\m\,\ldots,\nu)\,,\nn\\
\ll{newints}\\
&&\dfrac{d}{dt} s_m^{\pi}(\l,\m\,\ldots,\nu)=0\,,\nn
\en
The trace $\tr_{(m)}$ in (\ref{newints}) is taken over the whole space
$V^{(m)}$.

For any hamiltonians $I_{i,k}$ (\ref{int}) condition (\ref{eq1}) is
always fulfilled for the equivalent spectral parameters
\bq \l=\l_j=\l_{\pi(j)}\,,\qquad j=1,\ldots,n\,,\quad \forall\pi
\ll{point1}
\eq
that corresponds to a choice of another basis of ad-invariant
functions in the center $\gc(g)$  \cite{ks82} by using the
outer powers of matrix $L(\l)$
\bq
s_m(\l)=\dfrac1m\tr_{(m)} L^{(m,\pi)}(\l,\l,\ldots,\l)\,.\ll{sm}
\eq
Symmetric functions $s_m(\l)$ can be expressed in the symmetric
functions $\tau_m(\l)$ (\ref{tm}) according to Newton's
formulas \cite{ks82} and functions $s_m(\l)$ give rise to
integrals in the involution as before.

In additional, condition (\ref{eq1}) is  fulfilled for
the independent on spectral parameter matrix $A$ in
(\ref{laxmu})
\[
A_{i,m}(\mu)=\Phi^{(i)}_{\l}
\tr_1\left(r_{21}(\l,\mu)L_1^{m-1}(\l)\right)=const\neq 0\,.
\]
We observe that matrix $A_{i,m}(\mu)$ depends on spectral
parameter $\m$ via $r$-matrix only. Hence, the constant in
spectral sense matrix $A(\mu)$ and the corresponding special
hamiltonian are related to the singular points of the
$r$-matrix.

The  standard rational $r$-matrix for algebra $sl(n,\bC)$ is
equal to
\bq
r_{12}(\l,\m)=\dfrac{P_{12}}{\l-\m}\,.\ll{ratr}
\eq
We can choose the special linear functional in (\ref{lax2})
as residue at infinity
\bq
\Phi_\l(z)=-\left.{\rm Res}\right|_{\l=\infty}(\phi(\l)\cdot z)\,,
\ll{lf}\eq
such that the second matrix
\bq
A=\tr_1 \left[ P\cdot\lim_{\l\to\infty}\phi(\l)L^{m-1}_1(\l)\right]
\,.\ll{mrat}
\eq
is independent on spectral parameter.
Moreover, if some invariant polynomial $\tau_k(\l)$ (\ref{tm})
has a nontrivial residue at $\l=\infty$ on the phase space
\bq
H=-\left.{\rm Res}\right|_{\l=\infty}\,
\phi(\l)\tau_k(\l)\,,\ll{ssham}
\eq
which is chosen as a hamiltonian, the corresponding Lax matrix
$A$ (\ref{mrat}) do not equal to zero.  In this case the
singular point $\l=\infty$ of $r$-matrix (\ref{ratr}) is
associated with the superintegrable hamiltonian $H$
(\ref{ssham}) and with the new multivariable generating
functions of the integrals of motion (\ref{newints})
\ben
&&s_m^\pi(\l,\m,\ldots,\nu)=\dfrac1m\tr_{(m)} L^{(m,\pi)}(\l,\m,\ldots,\nu)
\,,\ll{smn}\\
&&\{H,s_m^\pi(\l,\m,\ldots,\nu)\}=0\,.\nn
\en
We may to lose the property of involution for these new
multivariable functions and the corresponding integrals of
motion
\bq
\{s_m^{\pi_m}(\l_1,\l_2,\ldots,\l_m),s_k^{\pi_k}
(\m_1,\m_2,\ldots,\m_k\}\neq 0\,.
\ll{linv}
\eq
These brackets are completely recovered by the polynomial
$r$-brackets for the matrices $L^{(m,\pi)}$, as an example see
(\ref{rp1}-\ref{rp2}).

The coefficients of the spectral curve of $L(\l)$ determined
by the characteristic equation
\bq
C(z,\l)=\det\left(zI+L(\l)\right)=0\,.\ll{curv1}
\eq
give rise to integrals in the involution only.
Complete set of integrals is determined by the generalized
spectral surfaces
\bq
C(z,\l,\m\,\ldots\,\nu)=
\det\left(zI+L^{(m,\pi)}(\l,\m\,\ldots\,\nu)\right)=0\,,
\qquad m\leq n\,,
\quad \forall \pi\,.\ll{curv2}
\eq
As an example, let us consider the $2\times 2$ Lax matrix
\bq
L(\l)=\left(\begin{array}{cc}h&e\\
f&-h\end{array}\right)(\l)\,,\ll{2lax}
\eq
in the two-dimensional auxiliary space with a suitable basis
\cite{ft87}
\[
P=\dfrac12(I+\sum \s_j\otimes\s_j)=
\left(\begin{array}{cccc}1&0&0&0\\
0&0&1&0\\ 0&1&0&0\\0&0&0&1\end{array}\right)\,,\]
where Pauli matrices $\s_j$ form an orthonormal basis of
$sl(2)$ in $\bC^2$.

Matrix $L(\l)$ has alone invariant polynomial
(\ref{sm})
\bq
s_2(\l)=\dfrac12 \tr \left[P L(\l)\otimes L(\l)\right]=
\det L(\l)=-\dfrac12 \tr L^2(\l)=-\dfrac12\tau_2(\l)\,,\ll{s2l}
\eq
and one second order complementary polynomial (\ref{smn})
\bq
s_2(\l,\m)=\dfrac12\tr\left[P L(\l)\otimes L(\m)\right]=
h(\l)h(\m)+\dfrac{e(\l)f(\m)+f(\l)e(\m)}2\,.
\ll{s2lm}
\eq
These polynomials may be the  generating functions of the
integrals of motion for some superintegrable hamiltonian
(compare with (\ref{adint})).

Next let us consider the Belavin $r$-matrices \cite{bel80}, which
are meromorphic solutions of the classical Yang-Baxter equation
such that
\ben
&&r_{12}(\l,\m)=r_{12}(\l-\mu)=-r_{21}(\mu-\l)\,,\ll{belr}\\
&&r_{12}(\l-\mu)=\dfrac{P_{12}}{\l-\mu}+O(1)\,,\qquad
P_{12}x\otimes y= y\otimes x\,.\nn
\en
Here $r_{ij}$ are rational, trigonometric or elliptic
matrix-function on spectral parameters \cite{bel80,ft87,rs87}.
These matrices have a pole at $\l=\mu$ with
a permutation operator $P_{12}$ (\ref{belr}) as residue.
In additional the rational function $r(\l-\m)$ of two variables
$\l$ and $\m$ has the special point at $\l=\infty$ in its domain
of definition.  In this point function $r(\l-\m)$ has a
distinct from zero residue, which is independent from second
spectral parameter $\m$.
The generally accepted elliptic $r$-matrices
\cite{bel80,ft87,rs87} have not such special points in its domains.
Nevertheless, the similar to the rational case
construction can be proposed as well.

The Lax equation (\ref{lax1}) and $r$-bracket (\ref{rpoi})
are covariant under the similar transformation
\bq
L\to U^{-1}LU\,,\qquad A\to U^{-1}AU\,,\qquad
r_{ij}\to U_1^{-1}U_2^{-1}r_{ij}U_1U_2\,,\ll{str}
\eq
where $U$ is a constant matrix on a phase space.  If the matrix
$U$ depends on spectral parameter, we can use such similar
transformations to construct the additional poles of
$r$-matrix.  In this case the multivariable generating
functions $s_m^\pi(\l,\m,\ldots,\nu)$ are changed
\bq
s_m^\pi(\l,\m,\ldots,\nu)\to s_m^\pi(\l,\m,\ldots,\nu)=
\dfrac1m\tr_{(m)} \left[
Z_\pi L(\l)\otimes L(\m)\ldots\otimes L(\nu)\right]\,,
\ll{chs}
\eq
here projector $P_\pi$ in (\ref{smn}) is substituted by matrix
\bq
Z_\pi=\left[
U(\l)\otimes U(\m)\ldots\otimes U(\nu)\right]^{-1}\cdot P_\pi
\cdot
U(\l)\otimes U(\m)\ldots\otimes U(\nu)\,.\nn
\eq
depending on spectral parameters. For the equivalent spectral
parameters functions $s_m(\l)$ are covariant under the similar
transformations.

As an example, here we study an elliptic $r$-matrix on the
twisted loop algebra $\SL(sl(2),\s)$. Let us consider the
period lattice $\Gamma=2K\bZ+2iK'\bZ$, where $K$ and $K'$ are
the standard elliptic integrals of the module $k\in[0,1]$, and
introduce the corresponding elliptic theta function
$\Theta_{ij}(\l,k)$ as in  \cite{skl86}.  In these notations
the standard elliptic $r$-matrix is
\bq
r(\l-\m)=\sum\limits^3_{k=1} w_k(\l-\m)\cdot
\sigma_k\otimes\sigma_k\,,\label{rell}
\eq
where $\sigma_k$ are the Pauli matrices and
\[
w_1(\l)=\frac{\Theta_{11}'(0,k)\Theta_{10}(\l,k)}
{\Theta_{10}(0,k)\Theta_{11}(\l,k)}
\,,\quad
w_2(\l)=\frac{\Theta_{11}'(0,k)
\Theta_{00}(\l,k)}{\Theta_{00}(0,k)\Theta_{11}(\l,k)}\,,
\quad
w_3(\l)=\frac{\Theta_{11}'(0,k)\Theta_{01}(\l,k)}
{\Theta_{01}(0,k)\Theta_{11}(\l,k)}\,.\]
Function $r(\l-\m)$ (\ref{rell}) is meromorphic in $\bC$ and has
simple poles at $\l=\m~{\rm mod}\Gamma$.

According to \cite{skl86} we introduce the similar $r$-matrix
\ben
&&\rho(\l,\m)=
U^{-1}_{12}(\l,\m,\l_\infty)\,r(\l-\m)\,
U_{12}(\l,\m,\l_\infty)\,,\nn\\
&&U_{12}(\l,\m,\l_\infty)=
U_1(\l-\l_\infty+K)U_2(\m-\l_\infty+K)\,.\nn
\en
with the following matrix $U(\l)$
\bq
U(\l)=\left(\begin{array}{cc}\Theta_{01}&\Theta_{00}\\
-\Theta_{00}&-\Theta_{01}\end{array}\right)\,
(\xi,\tilde{k})\,,\quad\l=\dfrac{2\xi}{1+k}\,,
\quad\tilde{k}=\dfrac{2\sqrt{k\,}}{1+k}\,,\ll{uell}
\eq
here $\l_\infty$ is an arbitrary point.  This $r$-matrix
$\rho(\l,\m)$ has been introduced for the purpose of separation
variables in \cite{skl86}.  More explicitly
\ben
\rho(\l,\m)&=&\sum_{j=1}^3
w_j(\l-\m,k)\cdot \sigma_j\otimes\sigma_j+\label{rhoxyz}\\
\nn\\
&+&
\left[\,w_3(\m-\l_\infty,k)-w_3(\l-\l_\infty,k)+w_0\,\right]
\cdot\sigma_3\otimes\sigma_3
\nn\\
\nn\\
&+&
iw(\m-\l_\infty,k)\cdot\sigma_3\otimes\sigma_2
-i w(\l-\l_\infty,k)\cdot \sigma_2\otimes\sigma_3\,;
\nn
\en
\[
w(\l,k)\equiv w_1(\l,k)=w_2(\l,k)=
\frac{\Theta'_{11}}{\Theta_{10}}
\frac{\Theta_{10}(\l,k)}{\Theta_{11}
(\l,k)}\,,\qquad
w_3(\l)=\frac{\Theta'_{11}
(\l,k)}{\Theta_{11}(\l,k)}\,,
\]
here $w_0$ is a some constant \cite{skl86}.  The poles of
$\rho(\l,\m)$ by the first spectral parameter $\l$ be at the
points
\ben
&\l=\m~{\rm mod}\,\Gamma\,,\qquad
&\left.{\rm Res}\right|_{\l=\mu} \rho(\l,\m)=P\,,\nn\\
&\l=\l_\infty~{\rm mod}\,\Gamma\,,\qquad
&\left.{\rm Res}\right|_{\l=\l_\infty} \rho(\l,\m)=Z=
(\sigma_3+\sigma_-+\sigma_+)\otimes\sigma_3\,.\nn
\en
So, if $L(\l)$ is an orbit of the new $r$-matrix $\rho(\l,\m)$,
such that the second matrix
\[A=\tr_1 \left[Z\cdot\lim_{\l\to\l_\infty}
\phi(\l)L(\l)\right]\neq 0\,,\]
is nontrivial constant in spectral sense,
the special hamiltonian
\[
H=\left.{\rm Res}\right|_{\l=\l_\infty}
\tr\left[\phi(\l)L^2(\l)\right]\]
is superintegrable.

In the next section we present some nontrivial examples of
$r$-matrix orbits associated with superintegrable systems for
the rational $r$-matrix.


\section{Examples}
\setcounter{equation}{0}
The above construction of superintegrable systems can be
applied to the Gaudin magnet, which was introduced in the
quantum mechanics \cite{gaud76}. The classical version turned
out to be useful example for developing a general
group-theoretic approach to integrable system \cite{ft87,rs87}.

We shall consider the rational Gaudin magnet related to $sl(N)$
algebra.  The model in question is defined on the $M$ coadjoint
orbits of $sl(N)^*$ in variables $X^{(m)}_{ij}$,
($m=1,\ldots,M$, $i,j=1,\ldots,N$). The corresponding
Lie-Poisson brackets are
\bq
\{X^{(m)}_{ij},X^{(n)}_{kl}\}=
\d_{mn}\left( X^{(m)}_{il}\d_{jk}- X^{(m)}_{kj}\d_{il}\right)\,.
\ll{sln}
\eq
The coadjoint orbits are fixed by values $t^{(m)}_k$ of the ad-invariant
functions on $sl(N)$
\bq
t^{(m)}_k=k^{-1} \tr (X^{(m)})^k\in\bC\,.\ll{orsln}
\eq
The Poisson bracket (\ref{sln}) is nondegenerate on the
manifold (\ref{orsln}) having dimension $n=MN(N-1)/2$ for the
case of generic orbit (all $t_i^{(m)}$ are distinct). In what
follows we assume that the orbit is generic.

Fixing some element $Z\in sl(N)$ as a residue at infinity we
consider the special Lax matrix $L(\l)\in \SL(\oplus sl(N))$
\bq
L(\l)=Z+\sum_{m=1}^M \dfrac{X^{(m)}}{\l-\d_m}\,,\ll{gaud}
\eq
where $\{\d_m\}$ is a set of $M$ arbitrary constants.
Matrix $L(\l)$ obeys the linear $r$-bracket (\ref{rpoi})
\cite{ft87,rs87} with the rational $r$-matrix (\ref{ratr}).

The basis elements $\tau_k$ (\ref{tm})
are meromorphic functions of $\l$
\[
\tau_k(\l)=\xi_k+\sum_{m=1}^M\sum_{j=1}^k
\dfrac{I_{m,k}^j}{(\l-\d_m)^j}\,,
\]
here $\xi_k=k^{-1} \tr Z^k$ and $I^k_{m,k}=t^{(m)}_k$
are fixed constants. Another residues
$I_{m,k}^j$ form a family of $n=MN(N-1)/2$ independent
integrals in the involution. It is immediately seen that
the special hamiltonians in this family
\[H^{(k)}=-\left.{\rm Res}\right|_{\l=\infty} \tau_k(\l)=
\sum_{m=1}^M \left.{\rm Res}\right|_{\l=\d_m} \tau_k(\l)
\,,\]
are nondegenerate functions on the generic coadjoint orbits of
$sl(N)^*$ and they are corresponded to the constant in spectral sense
second Lax matrix (\ref{lax2})
\[A^{(k)}=Z^{k-1}\,.\]
So, hamiltonians $H^{(k)}$ are superintegrable hamiltonians and
complete set of the integrals of motion can be generated by
(\ref{smn}).  As an example, additional integrals of evolution
may be constructed from the quantities
\bq
I_{m,k}^\pi={\rm Res}~ s_m^\pi(\l,\m,\ldots,\nu)\,.
\ll{addi}
\eq
Here Res means residue of some fixed order
$k=(k_1,k_2,\ldots,k_m)$, $k_j\leq K_j$ at the points
\[\l=\d_{j_1}\,,\quad \m=\d_{j_2}\,,\ldots\,,~\nu=\d_{j_m}\,,\]
which belong to the divisor of the poles of multivariable function
$s_m^\pi(\l,\m,\ldots,\nu)$
\[D=\{(\d_j,K_j)\,,j=1,\ldots,M\,,(\infty,1)\}\,.\]

As a second example, let us consider superintegrable natural
systems on $\bR^{2n}$ with the following hamiltonian
\bq
H=T+V=\sum a_{ij}p_ip_j+V(q_1,\ldots,q_n)\,,\qquad a_{ij}\in\bR\,,
\ll{nham}
\eq
where $\{p_j,q_j\}_{j=1}^n$ are canonical variables.  By $V=0$
in (\ref{nham}) this hamiltonian of geodesic motion is
superintegrable (\ref{fmot}).  In the classical mechanics the
geodesic motion on the Riemannian spaces of constant curvature
is isomorphic to the rational Gaudin magnet on the algebra
$\ggl=\oplus sl(2)$ \cite{kuz92}. Taking the known Lax matrix
$L(\l)$ in $\SL(sl(2),\l)$ associated to a free geodesic
motion we can get an infinite set of the Lax matrices
associated to the completely integrable potential systems
\cite{ts96b}. Construction of the new Lax equations
consists of the application of the outer automorphism of
infinite-dimensional representations of $sl(2)$ (\ref{mapp}) to
the loop algebras.  As a second step, we have to apply the
consequent projections of this mapping onto a suitable Poisson
subspaces of $r$-bracket (${\rm ad}_R^*$-invariant subspaces).
The superintegrable systems are related to the special
projections, which lead to a constant in spectral sense Lax
matrix $A(\l)$ (\ref{lax1}).

Let us consider the Lax matrix $L(\l)$ in $\SL(sl(2),\l)$
associated to geodesic motion.  Hereafter we propose that this
matrix $L(\l)$ in the form (\ref{2lax}) obeys the $r$-matrix
bracket (\ref{rpoi}) with the rational $r$-matrix (\ref{ratr}).
Following to the general procedure \cite{eekt94,krw95,ts96b} let
us introduce new Lax pairs
\ben
&&L'(\l)=L(\l)-\s_-\cdot\left[\phi(\l)e^{-1}(\l)\right]_{MN}\,,\nn\\
\ll{mapmn}\\
&&A'(\l)=A-\s_-\cdot\left[\phi(\l)e^{-2}(\l)\right]_{MN}=
\left(\begin{array}{cc}0&1\\
u_{MN}(\l)&0\end{array}\right)\,.\nn
\en
Here $\phi(\l)$ is a function on spectral parameter and
$[z]_{MN}$ means restriction of $z$ onto the ${\rm
ad}^*_R$-invariant Poisson subspace of the initial $r$-bracket
\cite{krw95,ts96b}.  For the rational $r$-matrix (\ref{ratr})
we can use the linear combinations of the following Laurent
projections
\bq
{[ z ]_{MN}}=\left[\sum_{k=-\infty}^{+\infty} z_k\l^k\,
\right]_{MN}\equiv
\sum_{k=-M}^{N} z_k\l^k\,,
\ll{cutmn}
\eq
or the Taylor projection by $M=0$.

The mappings (\ref{mapmn}) play the role of a dressing
procedure allowing to construct the Lax matrices $L'_{MN}(\l)$
for an infinite set of new integrable systems starting from the
single known Lax matrix $L(\l)$ associated to one integrable
model.  The Lax matrix $L'(\l)$ (\ref{mapmn}) obeys the linear
$r$-bracket (\ref{rpoi}), where constant $r_{ij}$-matrices
substituted by $r_{ij}'$-matrices depending on dynamical
variables
\bq r_{12}(\l,\mu)\to
r'_{12}=r_{12}
-\frac{\left( [\phi(\l) e^{-2}(\l)]_{MN}-
[\phi(\m) e^{-2}(\m)]_{MN}\,\right)}{(\l-\m)}\cdot\s_-\otimes\s_-
\,.\ll{dpr}
\eq

The Poisson map $\bR^{2n}\to\SL(\ggl)$ may be determined by
using a Newton form of the geodesic equations of motion
\cite{krw95}.  Letting
\bq
A=\left(\begin{array}{cc}0&1\\0&0\end{array}\right)
=\s_+\,,\ll{ain}\eq
which is independent on spectral parameter $\l$,
we observe that the Lax matrix $L(\l)$ (\ref{2lax}) is
recovered by the alone entry $e(\l)$ and by the hamiltonian
$H$ according to
\bq
L(\l)=\left(\begin{array}{cc}-{e_x}/2&e\\
-e_{xx}/2&{e_x}/2\end{array}\right)(\l),\quad
e_x=\{H,e\},\ll{sew}
\eq
In this case the generating functions $s_2(\l)$ and $s_2'(\l)$
obey the following equations of motion
\ben
\dfrac{d s_2(\l)}{dt}=0\,,\quad&\Rightarrow\quad
&\partial^3_x e(\l)=e_{xxx}=0\,,\nn\\
\ll{sewu}\\
\dfrac{d s_2'(\l)}{dt}=0\,,\quad&\Rightarrow\quad
&\left[\dfrac14\partial^3_x +u_{MN}(\l)\partial_x
+\dfrac12 u_{MN,x}(\l)\,\right]\cdot e(\l)=B_1[u_{MN}(\l)]\cdot e(\l)=0.
\nn
\en
Here $B_1[u_{MN}(\l)]$ is the Hamiltonian pencil operator for
the coupled KdV equation \cite{krw95}. The superintegrable
systems correspond to the special solution of (\ref{sewu}) by
$\partial u_{MN}(\l)/\partial\l=0$.  The integrate form of the
KdV recursion relations (\ref{sewu}) are
\ben
s_2(\l)&=&\dfrac{e\cdot e_{xx}}{2}-
\dfrac{e_x^2}{4}\,,\nn\\
\label{dob}\\
s_2'(\l)&=&\dfrac{e\cdot e_{xx}}{2}-
\dfrac{e_x^2}{4}+e^2\cdot u_{MN}(\l)\,.\nn
\en
A simple substitution for the entries of matrix $L(\l)$
\ben
e(\l)&=&{\cal B}^2\,,\qquad
h(\l)=-e_x/2=-{\cal B}{\cal B}_x\,,\label{subsb}\\
f(\l)&=&-e_{xx}/2=-{\cal B}_x^2-{\cal B}{\cal B}_{xx}\,,\nn
\en
turns determinants (\ref{dob}) into the form
\bq
s_2(\l)={\cal B}^3{\cal B}_{xx}\,,\qquad
s_2'(\l)={\cal B}^3{\cal B}_{xx}+{\cal B}^4
\left[\dfrac{\phi(\l)}{{\cal B}^4}\right]_{MN}\,,\label{dsub}
\eq
if we use an explicit formula for the potential $u_{MN}(\l)$.
These equations have the form of Newton's equations for the function
$\cal B$
\ben
{\cal B}_{xx}&=&s_2(\l){\cal B}^{-3}\,,\nn\\
\label{newton}\\
{\cal B}_{xx}&=&s'_2(\l){\cal B}^{-3}-{\cal B}
\left[\dfrac{f(\l)}{{\cal B}^4}\right]_{MN}\,,\nn
\en
If we assume that ${\cal B}=\sum_{j=0}^N q_{N-j}\, \l^j$ is a
polynomial, then its coefficients $q_j$ obey the Newton
equation of motion (\ref{newton}) with $s_2(\l)=\sum I_k\l^k$,
where $I_k$ are integrals of motion.  Here we reinterpret the
coefficients of $s_2(\l)$ and $s_2'(\l)$ in
(\ref{newton}) not as functions on the phase space,
but rather as integration constants.  In  variables $q_j$
mapping (\ref{mapmn}) affects only on the potential
($q$-dependent) part of the integrals of motion $I_k$.  The
kinetic (momentum dependent) part of $I_k$ remains unchanged.
So, the dressing mapping (\ref{mapmn}) allows us to get over
from a free motion on $\bR^{2n}$ to a potential motion on
$\bR^{2n}$.

To construct the multipole Lax equation we introduce an
appropriate completion $\SL_D(sl(2))$ of the standard loop
algebra associated to some fixed divisor of poles \cite{rs87}
\[D=\{(\d_j,l_j)\,,j=1,\ldots,M\,,(\infty,K)\}\,.\]
According by (\ref{sew}) the Poisson map
$\bR^{2n}\to\SL_D(sl(2))$ is completely defined by the alone
entry $e(\l)$ and by the hamiltonian $H$ related to
the linear functional $\Phi_\l$. We put
\ben
e(\l)&=&\sum_{i=1}^K e_i\l^i+\sum_{j=1}^M\sum_{k=1}^{l_j}
\dfrac{e_{jk}}{(\l-\d_j)^k}=
\dfrac{u_0\prod(\l-u_j)}{\prod(\l-\d_j)}\,,\nn\\
\ll{ine}\\
\Phi_\l(z)&=&
\left.{\rm Res}\right|_{\l=\infty} \left(\l^{-K}z\right)\,,\nn
\en
that self-consistent with (\ref{ain}) and (\ref{subsb}) \cite{krw95}.
Residues $e_i$ and $e_{jk}$ are the special functions of the
canonical coordinates $\{q_j\}_{j=1}^n$, and variables $u_j$
are curvilinear separated coordinates \cite{krw95}.  By using
function $\cal B$ residues $e_i$ and $e_{jk}$ are easily
restored in canonical variables $\{q_j\}_{j=1}^n$. In the
simple poles at $\l=\d_j$ function $e(\l)={\cal B}^2$
(\ref{ine}) has the following residues $e_{j1}=q_j^2$.
Parametrization of residues in the higher order poles is
discussed in \cite{krw95}. As an example,  we consider the
polynomial part of $e(\l)$ corresponding to a pole at infinity.
Let
\bq
{\cal B}(\l)=\sum_{j=0}^K q_{K-j}\l^j\,,\ll{sumb}
\eq
with $q_0=1$ and
\bq
e(\l)=\sum_{j=0}^K e_j\l^j\,,\qquad
h(\l)=\sum_{j=0}^K h_j\l^j\,,\qquad
f(\l)=\sum_{j=0}^K f_j\l^j\,,\ll{sets}
\eq
with $e_K=1$, $h_K=0$ and $f_K=0$
due to the definition of $A$ (\ref{ain}) and $\Phi_\l$
(\ref{ine}).  Taking (\ref{subsb}) and (\ref{ine}) into account
we get
\ben
e_j&=&\sum_{i=0}^{K-j} q_i\, q_{K-j-i}\,,\qquad
h_j=-\sum_{i=0}^{K-j} q_{i,x}\, q_{K-j-i}\,,\nn\\
\label{ent}\\
f_j&=&-\sum_{i=0}^{K-j} q_{i,x}\,q_{K-j-i,x}
-\sum_{i=0}^{K-j} q_{i,xx}\,q_{K-j-i}\,,\nn
\en
where $q_x=\{H,q\}$ and we used the Newton formulae for a
product of two sets.  Canonically conjugate to the coordinates
$q_j$ momenta $p_j$ can be derived from the $r$-matrix algebra
(\ref{rpoi}) \cite{krw95}. Some first polynomials are equal to
\ben
K=0\,,\quad e(\l)&=&1\,,\qquad h(\l)=0\,,\nn\\\nn\\
K=1\,,\quad e(\l)&=&\l+2q_1\,,\qquad -h(\l)=p_1\,,\nn\\\nn\\
K=2\,,\quad e(\l)&=&\l^2+2\l q_1+(2q_2 +q_1^2)\,,\nn\\\nn\\
-h(\l)&=&\l p_2+(p_1 +p_2q_1)\,,,\label{nrep}\\\nn\\
K=3\,,\quad e(\l)&=&\l^3+2\l^2 q_1+\l (2q_2 +q_1^2)
+2(q_3+q_1q_2)\,,\nn\\\nn\\
-h(\l)&=&\l^2 p_3+\l (p_2 +p_3q_1)
+(p_1+p_2q_1+p_3q_2)\,.\nn
\en
Notice that the kinetic part of the hamiltonian $H$ (\ref{nham})
has a nondiagonal form in these variables
\bq
T=\sum_{j=1}^{K}p_jp_{K+1-j}\,.\label{kin}
\eq

Now we turn to the superintegrable systems. For the Taylor
projections (M=0) (\ref{cutmn}) the mapping (\ref{mapmn})
preserves the property of superintegrability if and only if
$N\leq K$ according to (\ref{mapmn}) and (\ref{ine}).  Here $K$
is a highest order of a pole of the entry $e(\l)$ (\ref{ine})
at infinity and $N$ is a highest power in projection
(\ref{cutmn}).  By $N>K$ dynamical $r$-matrix (\ref{dpr}) has
the higher poles at $\l=\infty$ and the corresponding dynamical
systems are no longer the superintegrable systems.
All these superintegrable systems are related to the special
point of $r$-matrix and the associated second Lax matrix $A'$
(\ref{mapmn}) remains a constant in spectral sense under the
mapping (\ref{mapmn}).  All these superintegrable systems are
related to the special stationary flows of the KdV hierarchy
(\ref{sewu}) and genus of the associated spectral curve of
$L'(\l)$ determined by the characteristic equation
(\ref{curv1}) is no more than number of the degrees of freedom.

As an example, we present here several superintegrable systems.
Let the entry $e(\l)$ has the simple poles at $\l=\d_j$ and the
$K$ poles at $\l=infty$
\[e(\l)=P_K(\l)+\sum_{j=K+1}^{n}\dfrac{q_j^2}{\l-\d_j}\,,\]
where $P_K(\l)$ are polynomials given by (\ref{nrep}).
For the some first values of $K=N$ the corresponding
superintegrable potentials in (\ref{nham}) are
\ben
&K=0\,,\qquad&V=\sum_{j=1}^n \left(q_j^2+
\dfrac{\b_j}{q_j^2} \right)\,,\nn\\
&K=1\,,\qquad&V=4q_1^2+\sum_{j=2}^n \left(q_j^2+
\dfrac{\b_j}{q_j^2}\right)\,,\nn\\
&K=2\,,\qquad&V=4q_1^3-8q_1q_2+
\sum_{j=3}^n \left(q_j^2+\dfrac{\b_j}{q_j^2}\right)\,,\nn\\
&K=3\,,\qquad&V=-5q_1^4+12q_1^2q_2
-4q_2^2-8q_1q_3+
\sum_{j=4}^n \left(q_j^2+\dfrac{\b_j}{q_j^2}\right)\,.\nn
\en
If $N=K$, then the dressing mapping (\ref{mapmn}) has the $N+1$
arbitrary parameters given by the function
$\phi(\l)=\sum_{j=0}^N\a_j\l^j$.    These parameters are
related to the canonical shifts of variables $q_j\to q_j+\a_j$
and to the common rescaling $V\to\a_N V$ in the presented
potentials.   Additional integrals of motion may be constructed
by the rule (\ref{addi}) from the multivariable generating
function $s_2(\l,\m)$ (\ref{s2lm}). Of course,
for the oscillator all these integrals (\ref{addi})
coinside with known ones (\ref{tjk}).

We may construct the similar superintegrable systems with
rational potentials by using a general form of the entry
$e(\l)$ (\ref{ine}) with the higher order poles and applying
more general Laurent projection (\ref{cutmn}). The presented
method can be employed to construct superintegrable systems on
the other Riemannian spaces of constant curvature \cite{kuz92}.
The corresponding quantum systems may be obtained by canonical
quantization \cite{eekt94,ts96b}.

Taking into account the $r$-bracket (\ref{rpoi}) one can conclude
that the entries $e(\l)$ and $f(\l)$ could play the roles
similar to the standard creation and annihilation operators for
harmonic oscillator \cite{ks82,fld94}. By using the
similar transformation of matrices $L(\l)$ or $L'(\l)$ with
matrix $U$ (\ref{simtr}) at $\a_j=\a_k=1$ we can obtain the
symmetric representation of these matrices
\[
e(\l)=e(\l,a_1,a_2,\ldots,a_n)\,,\qquad f(\l)=e^+(\l)
\,,\qquad h(\l)=h^+(\l)\,.\]
In this symmetric representation of the Lax matrix the usual
method of spectrum-generating algebras \cite{bb88,mm79} is a
part of the standard Bethe ansatz \cite{ks82,fld94}. It should
be emphasized, that the algebraic Bethe ansatz is a
sufficiently universal procedure, which slightly depends from
particular system in question. It allows us to interpret
various concrete models as some representations of alone
generalized model, which is defined by its $r$-matrix only.

As a third example, let us consider the Calogero-Moser systems.
It is well known \cite{pe91}, that both the Toda models and the
Calogero-Moser models are obtained by hamiltonian reduction of the
geodesic motion on the cotangent bundle $T^*G$ of a Lie group $G$.
For the geodesic motion on symmetric spaces of zero curvature
the canonical $2$-form, the free hamiltonian and equations of
motion are equal to
\ben
&w=\tr(dy\wedge dx)\,,\qquad &H=\dfrac12 \tr(y^2)\,,\ll{hrat}\\
&\dot{x}=y\,,\qquad &\dot{y}=0\,.\nn
\en
For the geodesic motion on symmetric spaces of positive or
negative curvature these quantities read
\ben
&w=\tr(x^{-1}dy\wedge x^{-1}dx)\,,\qquad &H=\dfrac12 \tr(yx^{-1})^2\,,\ll{htr}\\
&\dot{x}=y\,,\qquad &\dot{y}=yx^{-1}y\,.\nn
\en
The hamiltonians (\ref{hrat}) and (\ref{htr}) have
the following sets of integrals in the involution
\[ I_k=\tr(y^k)\quad{\rm and}\quad
I_k=\tr(yx^{-1})^k\,.\]
The additional integrals - "projections
of angular momentum" (\ref{fint}) - are equal to
\bq
I_{jk}=\tr(qp^{j-1})\tr(p^k)-\tr(p^j)\tr(qp^{k-1})\,.\ll{anm}
\eq
Here $q=x$ and $p=y$ for the first equations of geodesic motion
(\ref{hrat}) and $q=\ln x$ or $q=\ln y$ with $p=yx^{-1}$ for
the second equations of geodesic motion (\ref{htr}).

In the reduction process the Lax matrices of the reduced system
are expressed in terms of $x$ by a formula of the type
$L=zxz^{-1}$, where $z$ is  some elements in $G$ \cite{pe91}.
For the geodesic motion (\ref{hrat}) associated to the Calogero
model with the rational potentials, the hamiltonian $H$
(\ref{hrat}) remains superintegrable and images of integrals
(\ref{anm}) are integrals of a reduced system \cite{w83}. For
the second geodesic motion (\ref{htr}) associated to the
Calogero-Moser model with the trigonometric potentials and to
the Toda model, the reduced hamiltonian do not commute with the
images of logarithmic additional integrals (\ref{anm}).
In the quantum mechanics  whole polynomial algebra of the
integrals of motion for the Calogero model has been introduced
in \cite{kuz96}.

Let us show as the superintegrable hamiltonian (\ref{hrat})
appears in the $r$-matrix formalism. For instance, consider the
Euler-Calogero-Moser system \cite{w83}. Introduce a set of
dynamical variables
$\{(q_j,p_j)\}_{j=1}^N$ and
$\{f_{ij}\}_{i,j=1}^N$ ($f_{ij}=-f_{ji}$) together with the
Poisson brackets
\ben
&&\{p_j,q_k\}=\d_{jk}\,,\\
&&\{f_{ij},f_{kl}\}=\dfrac12\left(
\d_{il}f_{jk}+\d_{ki}f_{lj}+\d_{jk}f_{il}+\d_{lj}f_{ki}\right)\,.\ll{oorb}
\en
In order to have a nondegenerate Poisson bracket it is assumed
that the variables $f_{ij}$ are restricted to a symplectic
submanifold of (\ref{oorb}). The hamiltonian and the Lax matrix
for the Euler-Calogero-Moser system \cite{w83} are given by
\ben
H&=&\dfrac12\sum_{j=1}^N p_j^2+
\dfrac12\sum^N_{i,j=1~ i\neq j}\,\dfrac{f_{ij}^2}{(q_i-q_j)^2}\,,\ll{hekm}\\
L(\l)&=&\sum_{j=1}^N p_j e_{jj}+
\sum^N_{i,j=1~i\neq j}\,\left(\dfrac1{q_i-q_j}+\dfrac1\l\right)
f_{ij}e_{ij}\,,\nn
\en
with the corresponding $r$-matrix in the form \cite{bab94}
\ben
r_{12}(\l,\m)&=&-\dfrac{\l}{\l^2-\m^2}\sum_{j=1}^N e_{jj}\otimes e_{jj}
\ll{rdin}\\
&-&\dfrac12\sum^N_{i,j=1~ i\neq j}\,
\left(\dfrac1{q_i-q_j}+\dfrac1{\l+\m}\right)e_{ij}\otimes e_{ij}\nn\\
&-&\dfrac12\sum^N_{i,j=1~i\neq j}\,
\left(\dfrac1{q_i-q_j}+\dfrac1{\l-\m}\right)e_{ij}\otimes e_{ji}\,,\nn
\en
where $(e_{ij})_{kl}=\d_{ik}\d_{jl}$.
In the reduction process this $r$-matrix inherits the singular
point $\l=\infty$ from the initial rational $r$-matrix. The
corresponding superintegrable hamiltonian (\ref{hekm}) may be
defined by (\ref{int}) with $\phi=1/2\cdot\l^{-1}$
\[
H=\Phi_\l\left[\tr L^2(\l)\right]=
\dfrac12 \left.{\rm Res}\right|_{\l=\infty}
\left[\l^{-1}\cdot\tr L^2(\l)\right]\,.\]
and the second Lax matrix is independent on spectral parameter
\[
A=\Phi_{\l}\tr_1\left[r_{21}(\l,\mu)L_1(\l)\right]=
\sum^N_{i,j=1~i\neq j}\,\dfrac{f_{ij}}{(q_i-q_j)^2}e_{ij}\,.\]
The higher flows with $\phi(\l)=1/k\cdot\l^{-k-1}$ in (\ref{int})
are superintegrable as well \cite{w83}.

\section{Conclusions}
\setcounter{equation}{0}
We have seen that superintegrable systems closed to
geodesic motion can be realizing as
isospectral flows on coadjoint orbits of loop algebras in
framework of the $r$-matrix formalism. All these systems with
rational potentials are associated to the special singular
point of $r$-matrix.

Another  classical superintegrable systems with an arbitrary
number of degrees of freedom are the Kepler problem
\cite{arn89,pe91} and the rational Calogero-Moser systems
\cite{w83}.  In the proposed scheme we can consider a free
geodesic motion on the momentum sphere and use the stereographic
projection with appropriate change of the time variable
to study the Kepler problem \cite{m70}.
However, this transformation could violate the $r$-bracket
(\ref{rpoi}) for the corresponding Lax matrix. It would be
interesting to construct the $2\times 2$ Lax matrix for the
Kepler problem and for the Kepler-like superintegrable
potentials listed in \cite{ev90}.

This research has been partially supported
by RFBR grant 96-0100537.

\end{document}